\documentclass[sigplan, nonacm, twocolumn]{acmart}

\renewcommand\footnotetextcopyrightpermission[1]{}
\author{John Stephan}
\affiliation{
  \institution{EPFL, Matter Labs}
  \city{Lausanne}
  \country{Switzerland}}
\authornote{Correspondence to John Stephan <\texttt{john.stephan@epfl.ch}>.}

\author{Matej Pavlovic}
\affiliation{
\institution{NEAR, Matter Labs}
\city{Zurich}
\country{Switzerland}}

\author{Antonio Locascio}
\affiliation{
\institution{Matter Labs}
\city{Paris}
\country{France}
}

\author{Benjamin Livshits}

\affiliation{
\institution{Imperial College, Matter Labs}
\city{London}
\country{England}
}

\AtBeginDocument{%
  }

\setcopyright{acmlicensed}
\copyrightyear{2024}
\acmYear{2024}
\acmDOI{XXXXXXX.XXXXXXX}

\usepackage{cleveref}
\usepackage{multirow}
\usepackage{bigstrut}
\usepackage{xspace}

\newcommand{\point}[1]{\par\smallskip\noindent\textbf{#1}. }
\newcommand{\centheader}[1]{\multicolumn{1}{c}{\textbf{#1}}}
\newcommand{\centheaderl}[1]{\multicolumn{1}{|c}{\textbf{#1}}}
\newcommand{\centheaderr}[1]{\multicolumn{1}{c|}{\textbf{#1}}}
\newrobustcmd{\tool}{\textsc{CrowdProve}\xspace}

\title{\tool: Community Proving for ZK Rollups}

\begin{document}
\sloppy
\begin{abstract}
Zero-Knowledge (ZK) rollups have become a popular solution for scaling blockchain systems, offering improved transaction throughput and reduced costs by aggregating Layer 2 transactions and submitting them as a single batch to a Layer 1 blockchain. However, the computational burden of generating validity proofs, a key feature of ZK rollups, presents significant challenges in terms of performance and decentralization. Current solutions rely on centralized infrastructure to handle the computational tasks, limiting the scalability and decentralization of rollup systems.

This paper proposes \tool, a prover orchestration layer for outsourcing computation to unreliable commodity hardware run by a broad community of small provers. We apply \tool to proving transaction batches for a popular ZK rollup. 
Through our experimental evaluation, we demonstrate that community proving can achieve performance comparable to, and in some cases better than, existing centralized deployments. Our results show that even systems utilizing modest hardware configurations can match the performance of centralized solutions, making community-based proof generation a viable and cost-effective alternative. 
\tool allows both the rollup operator and community participants to benefit: the operator reduces infrastructure costs by leveraging idle community hardware, while community provers are compensated for their contributions.

\end{abstract}

\maketitle

\section{Introduction}
\label{sec_intro}

Rollups have emerged as a popular solution for scaling blockchain systems.
A rollup %
is a transaction processing system built on top of another blockchain, such as Ethereum.
For this reason, rollups are often referred to as ``Layer 2''~(L2) systems in relation to the ``Layer 1''~(L1) blockchain they are built on.
While L1 blockchains typically offer strong guarantees of security, decentralization, and Sybil resistance,
they often sacrifice performance, leading to high transaction latency and low throughput.
These performance bottlenecks, combined with network congestion they tend to induce, result in high transaction fees on L1 blockchains.
All the above greatly hinders the systems' usability.

Rollups are a critical innovation in scaling blockchain systems that addresses those limitations.
The primary function of rollups is to aggregate multiple transactions into a single batch,
significantly reducing the amount of data stored and computation performed by the underlying L1 blockchain.
By executing transactions off-chain and submitting only succinct proofs or aggregated data to L1,
rollups substantially improve throughput and lower transaction fees.
While rollups inherit many of the properties of the underlying L1 blockchain system, such as Sybil resistance and state finality, they trade in some others, such as decentralization.

This paper focuses on zero-knowkedge ~(ZK) rollups, which compensate for the lacking decentralization through validity proofs~---~succinct pieces of data proving that rollup transactions have been processed correctly by the rollup operator. In particular, we are concerned with the computational process involved in generating validity proofs, and explore paths towards decentralizing this process.

Generating a validity proof is computationally expensive, generally requiring orders of magnitude more resources than transaction execution itself.
Proof generation involves a variety of computational tasks, each with unique hardware requirements.
Depending on the used proving system, some tasks are highly parallelizable and can benefit from GPU acceleration, while others require large memory capacities or significant CPU performance.
Moreover, some tasks produce intermediate results that are easy to verify, while others do not.
Orchestrating these tasks in an efficient way under practical constraints (such as availability and price of hardware) can easily become non-trivial.
All this makes the decentralization of proof generation a challenging problem.

Rather than aiming for full decentralization of the proof generation process, we propose a partial decentralization strategy.
In this approach, certain proving tasks are outsourced to community participants, while others remain under the control of the rollup operator.
The insight behind partial decentralization is that by selectively outsourcing tasks that are easy to parallelize or verify,
significant benefits can be realized without requiring a complete overhaul of the system architecture.

It is important to note that, strictly speaking, a system remains centralized as long as any of its critical components is centralized.
In our proposed solution, the control over the orchestration of individual tasks remains centralized within the rollup operator.
Our partial decentralization scheme, however, allows community participants to contribute computational resources in a way that is beneficial to both the participants and the rollup operator.
This step represents a meaningful move toward decentralization, although full decentralization will require additional innovations, such as decentralized orchestration (discussed further in Section~\ref{sec:discussion}).

\subsection{Our Approach in \tool}
\tool offers key advantages for all involved parties:
\begin{itemize}
    \item Community participants can monetize their idle hardware by contributing computational power to the proving process.
    \item Rollup operators benefit from reduced infrastructure costs by outsourcing computational tasks to the community,
    which is likely to offer computational resources at a lower price than the operator would otherwise pay for dedicated hardware or cloud-based solutions.
\end{itemize}
This paper presents a partially decentralized proving system we call~\tool.
We design it as an extension of the current proving system implementation of a popular ZK rollup.
This extension allows anyone with sufficiently performant hardware to participate in the proving process.
Importantly, the performance requirements can be met by consumer-grade hardware, such as regular personal computers.
Additionally, our design ensures that even if some provers behave maliciously or fail, the integrity of the system is not compromised.
While our implementation focuses on a particular PLONK-based system,
the principles we introduce are broadly applicable and can be extended to other proving systems based on related cryptographic techniques.

Our main insight is that, in this family of proof systems,
large parts of the proof generation process are highly parallelizable.
Moreover (and equally importantly) the individual parallel tasks can be executed by untrusted community participants.
In the case of ZK rollups, the final validity proof for a batch of transactions is recursively composed of thousands of smaller sub-proofs, each of which can be computed independently. These sub-proofs are, by their nature, easy to verify. As a result, incorrect or malicious submissions can be quickly detected and discarded with minimal impact on the system. This inherent parallelism allows the system to efficiently distribute the computational workload across many community provers, enabling significant scalability without sacrificing security.

The deployment of a proving system on community hardware, which is often less reliable and may have limited computational resources, does introduce constraints.
For instance, certain optimizations that would be feasible in a datacenter environment, such as those requiring high-bandwidth communication between machines,
become ineffective when relying on distributed community hardware connected over the internet.
Intuitively, one would expect that those constraints inevitably limit the performance of community-based proving.
\tool\ demonstrates, however this is not the case in practice.
Despite the limitations, our results demonstrate that community proving performs comparably with, and in some cases even outperforms, centralized datacenter deployments.

This is largely due to the embarrassingly parallel nature of the proving tasks, where the relatively small size of individual sub-tasks helps offset the impact of slower communication speeds and limited hardware capabilities.
Moreover, since \tool is designed to leverage consumer-grade hardware, anyone with a reasonably powerful machine can participate, increasing the number of provers.
This expanded pool of provers also compensates for slower proving times on less capable machines, allowing the system to maintain efficiency.
While a datacenter deployment could, in theory mimic community proving and perform better thanks to additional datacenter-specific optimizations,
such an approach (potentially involving tens of thousands of virtual machines) would most likely need to be rejected for economical reasons.

Additionally, our design strategically collocates key components of the proving process, minimizing the need for large data transfers between provers and the rollup operator.
As a result, network bandwidth ceases to be a significant bottleneck, and the system is primarily affected by minor latency, which is manageable in most practical scenarios.

\subsection{Contributions}
This paper makes the following key contributions:

\point{Community Proving}
We design and implement a community-based proving system called \tool that enables rollup operators to offload proving tasks to a distributed network of community provers.
At the core of \tool is the Job Distributor~(JD) mechanism, which acts as a coordinator between the rollup operator and community provers.
The JD is responsible for dynamically assigning proving tasks to available provers, tracking the progress of each task, verifying proofs upon submission, and ensuring that provers are appropriately compensated.
The JD mitigates potential delays by reassigning jobs whenever provers fail to respond within the designated time or submit invalid proofs.
This decentralized job distribution model improves the system's scalability and reduces its reliance on centralized cloud infrastructure, fostering greater community involvement in the rollup ecosystem.

\point{Byzantine Resilience}
\tool is designed to operate seamlessly under Byzantine failure assumptions.
Malicious or faulty community provers are effectively deterred, as they are not compensated for submitting incorrect proofs, and unreliable provers~---~those that take excessive time to compute proofs~---~do not cause system delays.
This is achieved through our Least-Recently-Processed (LRP) job reassignment mechanism, which ensures that jobs are promptly reassigned to other provers without impacting the overall performance of the system.

\point{Experimental Evaluation}
We conduct a comprehensive experimental evaluation of \tool, demonstrating that it can match, and in some cases, outperform the current centralized proving deployment used by ZK rollups. The two primary performance metrics for our evaluation are proving time and proving cost.
Proving time measures the total duration required to generate a valid proof for a batch of transactions, while proving cost accounts for the overall expenditure incurred by the rollup, including both infrastructure costs and payments to external provers. 
Our results reveal that even systems relying on consumer-grade hardware can achieve competitive performance, with the potential to outperform existing centralized systems in terms of both speed and cost-efficiency. We also analyze the trade-offs between decentralization and performance, highlighting how our community proving system can achieve similar or superior results while reducing infrastructure costs and encouraging broader participation.

\subsection{Paper Outline}
In Section~\ref{sec_background}, we provide a comprehensive background on rollups, with a focus on ZK rollups and the ZKsync system. We explain how ZK rollups achieve scalability through validity proofs and introduce the Boojum prover, which serves as the basis for our community proving system.

In Section~\ref{sec_system_design}, we present the design and architecture of our Community Proving system. We detail how the Job Distributor~(JD) interacts with both the core rollup system and community provers, and outline the steps of job assignment, proof generation, verification, and compensation.

In Section~\ref{sec_experimental_eval}, we describe our experimental evaluation setup and metrics, including proving time and proving cost. We present the empirical results of the performance of \tool in different hardware configurations, comparing the results to centralized proving systems like those used in the current deployment of ZKsync.

In Section~\ref{sec:discussion}, we discuss the broader implications of our work. This includes possible variants of community proving, potential paths toward full decentralization, the introduction of adaptive pricing models, and future research directions to enhance the security and scalability of the system.

Section~\ref{sec:related} provides a summary of related work. 

Finally, in Section~\ref{sec_conclusion}, we present our conclusions, summarizing the contributions of the paper and outlining potential areas for further development in community-driven ZK rollup proving systems.
\section{Background}\label{sec_background}
In this section, we provide the technical background necessary to understand the context and foundation of our work.
We begin by explaining the operation of rollups, with a particular focus on zero-knowledge (ZK) rollups.
We then introduce ZKsync, a leading implementation of ZK rollups, and describe how it leverages the Boojum prover to generate validity proofs efficiently.
This background will establish the groundwork for understanding the design and architecture of our proposed Community Proving system in Section~\ref{sec_system_design}.

\subsection{Rollups}
A rollup is a computer system extending an underlying L1 blockchain.
As introduced in \Cref{sec_intro}, the core objective of a rollup is to process transactions faster and more cost-effectively than the L1 chain.
The rollup system periodically
1) aggregates large numbers of transactions into batches,
2) executes them, and
3) submits the results of the execution as a single transaction (smart contract invocation) to the L1 chain.
This aggregation enables the rollup to amortize the cost of an L1 transaction over a large number of L2 transactions,
leading to lower fees per transaction, while retaining many of the properties of the underlying L1 blockchain system.

The faster and cheaper execution provided by rollups often comes, however, at the cost of decentralization.
The process of ordering and executing rollup transactions is typically managed by a centralized subsystem called the sequencer.
To ensure the correctness of this execution, despite a possibly faulty or even malicious rollup operator, two classes of approaches have been developed:
With \emph{fraud proofs} (used by optimistic rollups), anyone can re-execute the transactions to check for inconsistencies.
If anyone detects a misbehavior, the system penalizes the rollup operator.
\emph{Validity proofs} (used by ZK rollups) rely on zero-knowledge cryptographic techniques
to generate an easily verifiable proof that all transactions have been executed correctly.
The validity proof is then checked directly by a smart contract on the L1 chain, providing an immediate guarantee of the correctness of the execution.


A typical rollup architecture involves two entities: the sequencer and the prover.
The sequencer receives transactions from clients, orders them, aggregates them into batches, and executes them.%
\footnote{The execution of transactions may sometimes not be considered part of the sequencer's responsibility,
but, for simplicity, we consider transaction execution to also belong to the sequencer's tasks.}
The prover is tasked with generating the validity proofs that ensure the correctness of state transitions resulting from those transactions.
The proving process is computationally intensive and typically handled by centralized infrastructure.
However, as explored in this paper, there are opportunities to decentralize this process,
allowing public participants to assist in generating proofs, thereby reducing costs and enhancing system scalability.


%

\subsection{Proving Systems}
\label{sec_background_proving}
A proving system is a computational framework that receives batches of transactions from the sequencer and produces proofs that validate their correct execution. There are various implementations of proving systems, but many follow a common structure, particularly those that recursively compute a final batch proof by aggregating smaller partial proofs.
One example of such a framework is the Boojum proving system used in ZKsync.
This section outlines the general workflow of such a system.

Upon receiving a transaction batch from the sequencer, the system's witness generation component creates a set of \textit{witnesses}.
Each witness represents a part of the batch's execution, such as a fixed number of operations processed by the rollup’s virtual machine.
In practice, a witness is a relatively compact piece of data~---~typically in the order of kilobytes or megabytes~---~serving as input to the prover component.

A prover, which could be a community prover or a cloud-based machine, receives a witness and produces a corresponding proof. Since the computation of proofs for different witnesses is independent, multiple instances of the prover can operate in parallel, often on separate machines. Once several proofs have been generated, another component of the proving system aggregates them to produce intermediate witnesses, continuing the recursive process until the final batch proof is assembled.

The process of computing a proof from a witness consists of multiple stages and produces intermediate results.
Some of these computation stages can benefit from hardware acceleration, such as GPU-based computation, and they could also lead to intermediate results being much larger than the original witness or final proof.
However, the final ZK proof, by definition, remains compact and easy to verify. The task of generating these proofs from witnesses, known as circuit proving, represents a significant portion of the total computational effort required to produce the final batch proof.

We define each unit of work performed by the proving system as a job.
The rollup operator maintains a database that tracks the status of all jobs, categorizing them as ``waiting for execution,'' ``running,'' or ``completed''.
When the system receives a new transaction batch, it initializes the database with jobs corresponding to all the witnesses generated from that batch. These jobs are then assigned to available provers, who compute the required proofs and update the database with their results, continuing this process until the final proof for the batch is complete.

Typically, the initial set of jobs is very large~---~often numbering in the (tens of) thousands~---~offering a high degree of parallelization potential.
However, fully leveraging this parallelism is a non-trivial task.
The computational load arrives in bursts with each new transaction batch, requiring a large number of provers simultaneously. Setting up a prover machine involves significant overhead, as hundreds of gigabytes of data may need to be downloaded or generated before the machine can begin processing.
This overhead makes it impractical to dynamically scale up or down by commissioning and decommissioning machines in a datacenter environment to match fluctuating demand.
A practical approach, therefore, is to maintain a smaller and more consistent pool of prover machines that take longer to process jobs, but spend less time waiting for new batches.
While this approach can reduce infrastructure costs, it comes at the expense of increased overall proof generation latency.
Finding the right balance between resource utilization and performance is a key challenge in designing efficient proving systems.

\section{\tool for Community Proving}
\label{sec_system_design}
The core idea behind our system, which we refer to as \tool, is to offload computational tasks involved in generating proofs to the public, allowing individuals or entities with spare computing power (e.g., GPUs, CPUs, consumer-grade laptops) to participate in the proving process by executing proving jobs.
These participants, known as community provers, may be incentivized to contribute through monetary compensation.
By leveraging the distributed resources of the community, the cost of generating ZK rollup proofs can be significantly reduced, while simultaneously fostering greater engagement and decentralization.

\subsection{System Architecture}
\tool differs from traditional cloud-based proving systems in three key ways.
First, we decouple the prover component (which we call community prover module) from the core system and make it publicly available for anyone to download and run.
This allows community participants to become active provers, contributing their computational resources.
Second, we introduce the Job Distributor (JD), a new component of the proving system that acts as a proxy between the proving system and the community provers. It distributes jobs stored in the proving system's database to the community provers.
Third, unlike conventional rollup systems that rely on core-system timeouts for job reassignment, \tool integrates a Least-Recently-Processed (LRP) job reassignment mechanism directly at the JD level.
This mechanism ensures better system performance by mitigating delays and improving the overall efficiency of the proving process.

\begin{figure}[tb]
    \centering
    \includegraphics[width=0.45\textwidth]{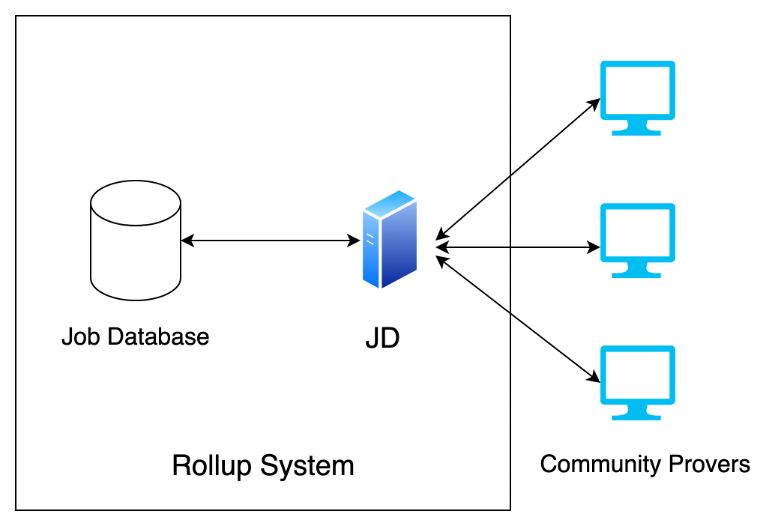}
    \caption{System Architecture of \tool}
    \label{fig_system_architecture}
\end{figure}

The system architecture of \tool, depicted in Figure~\ref{fig_system_architecture}, consists of three primary components: the core rollup system, the JD (managed by the rollup operator), and the community provers. This architecture is designed to be both flexible and scalable, supporting participation from a broad range of machines and reducing reliance on expensive cloud infrastructure.
In the following subsections, we provide a detailed overview of the architecture components in Sections~\ref{sec_comm_prover_module} and~\ref{sec_jd}, and explain the LRP mechanism in Section~\ref{sec_lrp}.

\subsubsection{Community Prover Module}\label{sec_comm_prover_module}
This module is a key component that enables public participation in the proving process.
It consists of software that community members can download and run on their own machines, allowing them to contribute computational resources to the system.
Once installed, the module handles the execution of individual proving jobs, taking witness data as input and generating cryptographic proofs, which are compact and easy to verify, as output.
Unlike a centralized datacenter deployment, where large intermediate results might be exchanged between high-performance machines, the Community Prover module is designed to minimize bandwidth usage, with both its input and output kept small (see \Cref{sec_background_proving}).
Additionally, certain parts of the proving process are amenable to hardware acceleration, allowing users with GPUs to perform tasks more efficiently.
The module communicates with the system’s Job Distributor (presented below) to request and submit jobs, ensuring that public participants can seamlessly integrate into the decentralized proving system.

\subsubsection{Job Distributor}\label{sec_jd}
The Job Distributor (JD) is the coordinating component that connects the core rollup system with community provers.
Rather than performing the computational tasks directly, the JD acts as a proxy, assigning proving jobs to available community provers and tracking the status of these jobs throughout the process.
The JD plays a crucial role in managing the flow of jobs, ensuring that tasks are distributed efficiently and completed in a timely manner.
When a prover requests a job, the JD fetches the next available task from the rollup system’s database and forwards it to the prover.
The JD also interfaces with the core system to verify proofs and update job statuses, ultimately compensating provers once their tasks have been successfully verified.
Additionally, if a job is not completed within the expected timeframe, the JD is responsible for reassigning the task to another prover via the LRP mechanism discussed below, ensuring that the system remains reliable and resilient against delays or failures.
The JD is designed to handle interactions with multiple community provers simultaneously, making it scalable and adaptable to a large, distributed network of participants.

\subsubsection{Least-Recently-Processed (LRP) Job Assignment Mechanism}\label{sec_lrp}
Traditional rollup systems often use a timeout mechanism to handle job reassignment. For example, in ZKsync~\cite{zksync}, the core rollup system triggers a timeout if a proving job is not completed within a set period (currently 10 minutes). However, this centralized timeout-based approach can be inefficient, particularly when dealing with Byzantine provers or malicious actors who may repeatedly claim jobs but fail to complete them, effectively limiting the contribution of honest participants to the system.

In contrast, \tool introduces the Least-Recently-Processed (LRP) job assignment mechanism directly at the JD level, eliminating the need for core rollup timeouts and improving the overall efficiency of the system. The JD maintains a local queue of pending jobs — those assigned but not yet completed. When a community prover requests a new job, the JD assigns it the least recently processed job from the queue (i.e., the job at the front of the queue), even if it has already been assigned to another prover. This ensures that resources are not wasted on faulty or malicious provers, while keeping the system's throughput high by continuously reassigning jobs.

This LRP mechanism also addresses potential Byzantine attacks, where malicious provers request and hoard jobs to disrupt the system. By reassigning jobs from the pending queue to honest provers, the system ensures that correct provers can continue working without waiting for the timeout period to expire. This strategy effectively improves resource utilization and mitigates delays, ensuring that the system remains efficient even in the presence of faulty or malicious provers. For further details, refer to the Trust and Reliability discussion in Section~\ref{sec_comm_prov_challenges}.

\subsection{Proving Process}
The proving process in \tool follows a structured, step-by-step workflow to ensure that jobs are efficiently assigned, proofs are correctly computed and verified, and provers are fairly compensated.
Below, we describe each stage of the process in detail:

\point{1) Job Request}
A community prover, ready to perform proving tasks, polls the JD to request jobs by calling the \texttt{get\_job} API of the JD.


\point{2) Job Fetch and Assignment}
The JD retrieves the next available proving job from the core rollup system’s database, increments its local \textit{request\_id} counter, and assigns the job (along with the \textit{request\_id}) to the requesting prover.
Once assigned, the JD marks the job as "running" in the core rollup system’s database to indicate that a prover is currently handling it.
If no new jobs are available from the rollup system~---~a scenario that can occur due to the bursty nature of transaction batch generation~----~the JD assigns the job at the head of its local queue of pending jobs to the prover (see \Cref{sec_lrp}).~\footnote{If no jobs are available in either the rollup system or the JD’s local queue, the prover is notified to try again later.}
The \textit{request\_id} serves as a unique identifier for each job request, allowing the JD to track multiple instances of the same job (with the same \textit{job\_id}) — especially in cases where a job is assigned to multiple provers concurrently.
This redundancy is necessary when the initial prover either submits an incorrect proof or takes too long to respond.
If the job was fetched from the rollup system, the JD adds it to the tail of the queue; if it’s already a pending job, the JD reassigns it and moves it to the back of the queue, making it the most recently processed job.
After assigning the job, the JD updates its internal hashmap, where the \textit{job\_id} serves as the key and is mapped to the corresponding \textit{request\_id}.
This hashmap is crucial for tracking multiple \textit{request\_id}s associated with the same \textit{job\_id} and is ultimately used during the proof verification process (described in step 5).

\point{3) Proof Computation}
Once the prover receives the job, it begins computing the proof. This involves several computationally intensive steps, depending on the specifics of the ZK proof system in use.

\point{4) Proof Submission}
After computing the proof, the prover sends the result back to the JD, together with the corresponding (\textit{job\_id}, \textit{request\_id}) pair, by calling the \texttt{submit\_result} API of the JD.

\point{5) Proof Verification}
Upon receiving the proof, the JD first verifies whether the submitted (\textit{job\_id}, \textit{request\_id}) pair is present in its local hashmap of active jobs.
If the pair is not found, the JD treats the submission as a potential fraudulent attempt and immediately discards it.
If the \textit{job\_id} and \textit{request\_id} are valid, the JD initiates the proof verification process to assess the correctness of the submitted proof.

\point{6) Finalization and Compensation}
Once the proof is verified successfully, the JD updates the core system’s database to mark the job as complete.
It then removes the job from its local queue of pending jobs and deletes the \textit{job\_id} along with all associated \textit{request\_id}s from the hashmap.
At this point, the JD compensates the community prover that submitted the first valid proof, effectively finalizing the job.~\footnote{In this work, we adopt a fixed-rate compensation model. For further details on this model, please refer to Section~\ref{fixed_rate_compensation}.}
By deleting the key associated with the job upon successful proof verification, the system effectively prevents any future proofs or responses from other provers — who may have been concurrently assigned the same job — from being processed or rewarded.

\point{Job Polling}
Provers must continuously poll the JD for new jobs if they wish to receive more tasks. We designed the JD to be as stateless as possible, avoiding the need for a persistent list of registered provers. For ease of implementation and efficiency, provers simply poll the JD (whose public IP address is assumed to be known to all interested community provers), and the JD replies directly with jobs when available. This design minimizes the overhead of maintaining a continuously updated registry and ensures that provers only receive jobs when they are ready. The polling mechanism also simplifies the JD, making it easier to scale and maintain.

\subsection{Benefits of \tool}
The key benefits of \tool are two-fold.

\point{Cost Reduction}
By outsourcing the proving process to community provers, the rollup operator can reduce reliance on costly cloud infrastructure. Community participants are typically willing to offer their computational resources at a lower price than renting cloud servers, resulting in overall cost savings for the rollup operator.

\point{Community Engagement and Incentives}
Community Proving introduces an opportunity for broader participation in the rollup ecosystem. By allowing individuals or organizations to contribute their hardware in exchange for compensation, this system creates a more decentralized and community-driven proving network.

\subsection{Challenges of Community Proving and \tool's Solutions}\label{sec_comm_prov_challenges}
While decentralizing computational tasks through community proving introduces several benefits, it also presents challenges, primarily around trust, reliability, and scalability. However, our system design of \tool inherently mitigates these challenges.

\point{Trust} 
Public provers cannot be fully trusted to always compute proofs correctly. While most proofs are expected to be valid, \tool handles occasional incorrect submissions gracefully by detecting incorrectly submitted jobs and reassigning the jobs without introducing major delays.
The JD verifies all incoming proofs before accepting them into the core system, eliminating the need to trust individual provers.
The verification of proofs does not involve full proof recomputation, but rather checks that the submitted proofs meet certain cryptographic conditions, minimizing computational overhead.
Additionally, economic incentives (e.g., withholding compensation for incorrect results) help deter malicious actors.
If proof verification fails, the JD does not mark the job as complete.
The job is not abandoned, however; instead, it remains in the JD's local queue of pending jobs, where it can be reassigned to another prover at a later time.
This approach allows the system to handle incorrect or malicious submissions without significantly impacting overall performance or increasing costs.

\point{Reliability}
Given the diverse nature of community provers, maintaining consistent performance can be challenging. Some provers may fail to complete jobs in a timely manner, as they may not have the necessary resources to handle certain types of proofs.
\tool addresses this by allowing the JD to reassign pending jobs to other provers when necessary (as discussed in~\Cref{sec_lrp} and under Trust).
In such cases as well, the pending job remains in the JD's local queue of active jobs, ensuring it can be reassigned to another prover at a later time, thereby minimizing delays and ensuring the system remains reliable and efficient.

\point{Job Hogging and DoS Prevention}
One potential threat to the system is a denial-of-service (DoS) attack carried out by malicious provers.
In such an attack, a prover could continuously request jobs without any intention of completing them, thereby monopolizing job assignments and disrupting the overall system efficiency.
By hogging available jobs, the malicious prover could prevent legitimate provers from receiving tasks, effectively stalling the proving process or causing significant delays in the system.
\tool mitigates this threat through the Least-Recently-Processed (LRP) job reassignment mechanism. Even if a Byzantine prover monopolizes all available jobs, these jobs remain in the JD's local queue of pending tasks, ensuring they are still available for reassignment to other, potentially honest, provers.
This ensures that the proving process continues without excessive delays.
Additionally, a rate-limiting mechanism could be introduced as an additional safeguard, restricting the number of jobs that can be assigned to a single prover within a specific time window. This would further ensure a fair distribution of jobs across the network of provers, preventing any single entity from monopolizing resources and incentivizing honest participation. By combining LRP with this rate-limiting mechanism, \tool remains robust and efficient even in the face of malicious behavior.
   
\point{Scalability}
As the number of community provers increases, the system must efficiently manage job distribution, communication, and proof verification to avoid performance bottlenecks.
One critical factor in decentralized proving is managing network bandwidth, as different stages of the proving process produce outputs of varying sizes (some of which could be large). To address this, \tool is designed to selectively outsource tasks in a way that keeps both input and output data manageable, minimizing the impact on bandwidth.
The main data exchanged between the provers and the JD are proving jobs and the resulting proofs, both of which are relatively small in size.

On average, a proof is approximately~742~KB, while most proving jobs are around~470~KB.
As a result, network bandwidth between provers and the core system is not a limiting factor, and the system’s performance is then primarily constrained by the computational time required to generate a proof (see~\Cref{sec_experimental_eval}).
Additionally, proof verification is computationally inexpensive, typically requiring only tens of milliseconds on a single-core machine.
By verifying cryptographic conditions rather than recomputing the entire proof, the JD can handle high volumes of proofs, even with thousands of active provers.
The simplicity of the JD’s design, which primarily involves forwarding jobs to provers, storing pending jobs, and verifying proofs upon receipt, enables it to minimize end-to-end latency. This design ensures that the JD can quickly distribute jobs and verify proofs, even in large-scale systems.
\section{Experimental Evaluation}\label{sec_experimental_eval}
In this section, we present the results of our experimental evaluation of the Community Proving system. Our goal is to quantify the performance of \tool in terms of proving time and proving cost across different hardware configurations.
Rather than competing with existing parallelization methods in the literature, \tool leverages these approaches to enable ZK proving within a community-driven context. This evaluation therefore focuses on assessing the feasibility of community proving as a computationally and cost-effective solution for ZK rollups.

\subsection{Experimental Setup}
To explore the effects of hardware capacity on system performance, we evaluate three distinct prover configurations:

\begin{enumerate}
    \item CPUs: Standard cloud-based machines with varying CPU core counts (8-core and 16-core configurations).
    \item MacBooks: Apple M3 Max machines with a mix of high-performance and low-power cores.
    \item GPUs: Machines with high-performance GPU configurations for accelerated proving.
\end{enumerate}

The experiments are designed to assess how each hardware configuration affects the total proving time for a batch of transactions.
Our tests simulate different scales by varying the number of provers, starting from a small pool of provers to a large-scale distributed network.
We focus on two primary metrics to analyze system performance:

\point{Proving Time} Measured as the duration required to prove a full batch of transactions, from the assignment of the first proving job from the batch to a community prover to the submission of the final proof to the JD.
This metric is crucial for evaluating the efficiency of our community-driven system in comparison to centralized, cloud-based approaches.
    
\point{Proving Cost} Calculated as the total expenditure associated with proving a batch of transactions, including mainly the compensation paid to community provers.
For comparison, we also consider the cost of renting cloud infrastructure under the current ZKsync configuration.

\medskip
We also assessed network delays in terms of job and proof exchanges between community provers and the JD, which averaged around 10 ms.
Given this minimal delay relative to the proving computational times reported in Table~\ref{table:proving_times_hardware}, we disregard network latencies in our evaluation to focus on the core performance of \tool.
Finally, since Section~\ref{sec_system_design} already addresses \tool’s ability to mitigate Byzantine attacks and malicious behavior, our experiments assume honest provers to study the feasibility of community proving in terms of proving cost and time.

\subsection{Empirical Results}
We conducted multiple experimental runs, varying the number of provers across different hardware configurations.
The results are presented below.

\begin{table*}[t]
\centering
\begin{tabular}{|c|c|c|c|c|}
\hline
\textbf{Hardware} & 8-core CPU & 16-core CPU & MacBook & GPU \\ \hline
\textbf{Job Proving Time (seconds)} & 97.6 & 67.7 & 48.5 & 14.4 \\ \hline
\end{tabular}
\caption{Proving time per job in seconds for different hardware configurations.
We consider two E2-standard CPU machines with 32 GB of RAM, one with 8 cores and another with 16 cores.
Furthermore, we also use a 4-core L4 GPU with 32 GB of RAM, as well as an Apple M3 Max MacBook with 36 GB of RAM and 14 cores (10 high performance, 4 low).}
\label{table:proving_times_hardware}
\end{table*}

\begin{figure}[t]
\centering
\includegraphics[width=0.49\textwidth]{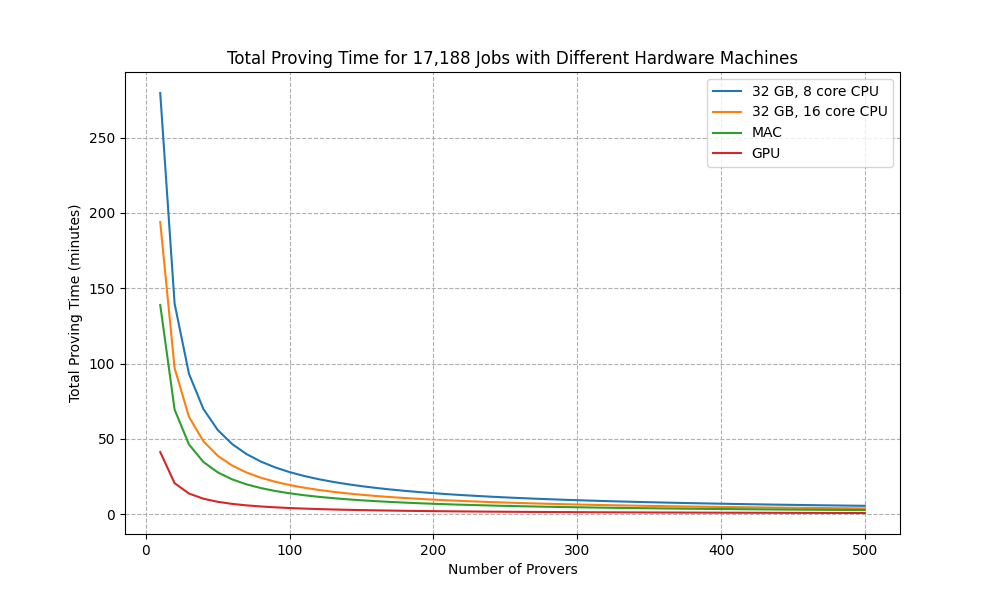}\\
\includegraphics[width=0.49\textwidth]{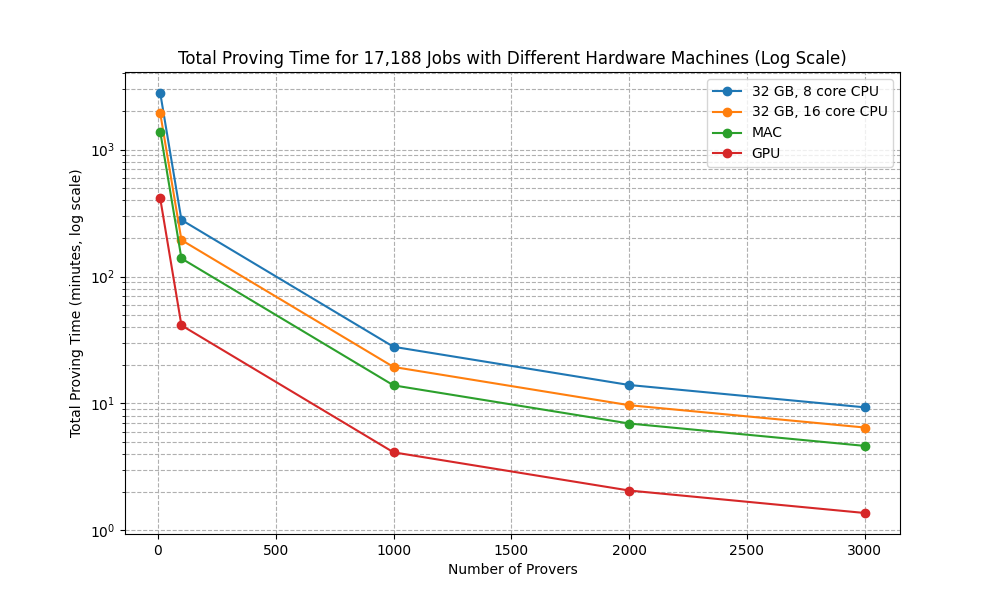}
\caption{Proving time in minutes for 17,188 jobs when varying the number of provers under different hardware configurations. \textit{Left}: linear scale, \textit{right}: logarithmic scale for better clarity on lower values.}
\label{fig:proving_times}
\end{figure}

\point{Impact of Hardware Configurations}
First, we present in Table~\ref{table:proving_times_hardware} the proving times per job for the different types of hardware we consider.
Based on the average proving times (per job) reported in Table~\ref{table:proving_times_hardware}, we extrapolated the total proving time for 17,188 jobs under perfect parallelization assumptions.
This corresponds to batch number 491452, comprised of 17,188 jobs of Aggregation Round 0, in the current ZKsync deployment.
Table~\ref{table:total_proving_times_minutes} presents the total proving time for 17,188 jobs across different hardware configurations and for different numbers of community provers, assuming ideal job distribution among the provers.
Figure~\ref{fig:proving_times} depicts the proving times on linear and log scales.
From these results, we can infer several key trends.
First, as the number of provers increases, the total proving time decreases across all hardware configurations (CPU, MacBook, GPU), as expected.
Second, the configuration of provers significantly impacts the overall proving time.
GPU machines exhibit the best performance due to their parallelization capabilities, making them ideal for reducing proving time.
MacBooks, which are consumer-grade products, also perform well, leveraging a combination of high-performance cores and efficient memory management.
CPU machines, though less efficient, can still achieve comparable performance at scale, especially when the number of community provers increases.
With 3,000 community CPU provers with 8 cores, \tool can prove a real batch in under 10 minutes, which is approximately $4 \times$ faster than the performance of the current centralized deployment of ZKsync.
Note that with at least 723 such community provers (a feasible number for distributed networks), \tool starts outperforming ZKsync's deployment system in terms of batch proving time.
Even fewer provers are needed when using the consumer-grade MacBook in our evaluation; just 359 such devices reach the breakeven point.
These results underscore the performance potential of a community-driven proving approach, demonstrating that \tool can outperform current centralized deployments with a reasonably low number of provers.

\begin{table*}[t]
\centering
\begin{tabular}{|r|r|r|r|r|}
\hline
\centheaderl{Provers} & \centheader{32 GB, 8 core CPU} & \centheader{32 GB, 16 core CPU} & \centheader{MacBook} & \centheaderr{GPU} \bigstrut \\ \hline\hline
10   & 2,795.91 & 1,939.38 & 1,389.37 & 412.52 \\ \hline
100  & 279.59   & 193.94   & 138.94   & 41.26  \\ \hline
1,000 & 27.96    & 19.39    & 13.89    & 4.12   \\ \hline
2,000 & 13.98    & 9.69     & 6.95     & 2.06   \\ \hline
3,000 & 9.32     & 6.46     & 4.63     & 1.37   \\ \hline
\hline
Breakeven - Provers & 723 & 502 & 359 & 107\\
\hline
\end{tabular}
\caption{Total proving time in minutes for 17,188 jobs under different hardware configurations.
It currently takes~38.70 minutes to prove these jobs in the ZKsync system.
The breakeven point shows the number of each type of prover needed to match the performance of the current deployment of ZKsync.}
\label{table:total_proving_times_minutes}
\end{table*}

\begin{table}[ht!]
\centering
\begin{tabular}{|l|r|r|r|r|}
\hline
\textbf{Number of Machines} & 5 & 10 & 15 & 20 \\ \hline
\textbf{Proving Time (min)} & 37.11 & 19.18 & 12.98 & 9.12 \\ \hline
\end{tabular}
\caption{Total proving time in minutes for~100 jobs of Circuit ID~1 and Aggregation Round~0 when varying the number of (8 CPU,~32 GB RAM) machines (provers).}
\label{table:total_proving_times_minutes_100_jobs}
\end{table}

\begin{figure}[tb]
    \centering
    \includegraphics[width=0.5\textwidth]{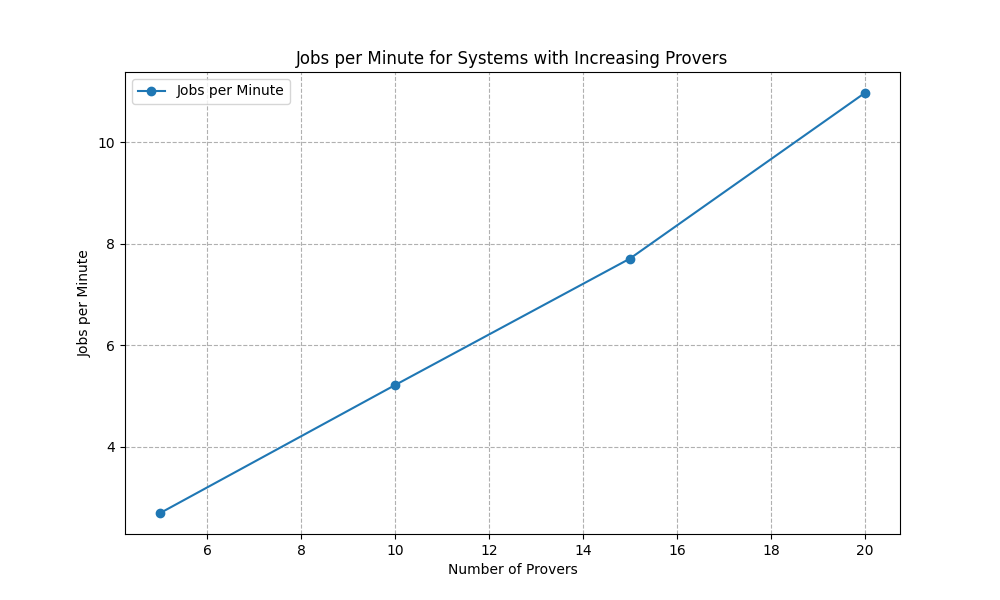}
    \caption{Total number of jobs executed per minute, when proving~100 jobs of Circuit ID 1 and Aggregation Round 0 across varying number of (8 CPU,~32 GB RAM)-provers.}
    \label{fig:proving_times_varying_machines}
\end{figure}

\point{Empirical scalability of \tool}
To validate the scalability of our assumptions in Table~\ref{table:total_proving_times_minutes}, we deployed a distributed system with~5,~10,~15, and~20 machines and measured the total proving time.
For simplicity, we consider (8-core,~32 GB RAM) CPU machines, and a total of~100 proving jobs.
Table~\ref{table:total_proving_times_minutes_100_jobs} and Figure~\ref{fig:proving_times_varying_machines} summarize these results.
As expected, the number of jobs proven per minute increases almost linearly with the number of machines, confirming the scalability of our system (as well as the  ideal parallelization and job assignment assumptions made in Table~\ref{table:total_proving_times_minutes}), particularly in configurations with CPUs.
These results validate the assumptions made in Table~\ref{table:total_proving_times_minutes}, showing that \tool can scale efficiently with the number of provers.

\point{Proving Cost Under Fixed Compensation}
In this evaluation, we assume a fixed compensation model where provers are compensated equally for each proving job they execute.
Matter Labs currently estimates the cost per prover job at approximately 0.0012 USD (0.12 cents).
By compensating community provers with a lower amount, we can effectively reduce costs while simultaneously decreasing batch proving time.
For example, in a system with at least~360 MacBook provers, where each prover is compensated 0.06 cents per proof, we match the performance of the current system (in terms of proving time, see Table~\ref{table:total_proving_times_minutes}) while reducing the cost by half.
Note that this payout is reasonable since such community provers would be able to execute on average~1,781 proving jobs daily, which results in being paid~1.07 USD per day.
This demonstrates the cost-effectiveness of our approach under a fixed compensation model.



\point{Trade-offs and Scalability}
Our experiments demonstrate that \tool can scale efficiently across various hardware configurations.
The trade-offs between proving time and proving cost are largely determined by the configuration of provers and the compensation model used.
Mixed configurations, combining limited machines (CPUs) with more powerful machines (GPUs), would provide a balance between performance and cost, allowing operators to optimize for both depending on their specific requirements.

By carefully selecting prover configurations and applying the appropriate compensation models, \tool achieves significant improvements in proving efficiency and cost reduction compared to traditional cloud-based proving systems like the current implementation of ZKsync.
These findings suggest that community proving, with the right balance of hardware and economic incentives, can support the broader adoption of ZK rollups in decentralized systems.

\point{Conclusion}
Our experimental evaluation provides compelling evidence that decentralized proving, using a community of provers with diverse hardware configurations, can achieve competitive performance in terms of both proving time and cost.
By optimizing prover configurations and adopting dynamic compensation models, community proving offers a viable alternative to centralized, cloud-based proving approaches, contributing to the scalability and cost-effectiveness of ZK Rollups.
\section{Discussion}
\label{sec:discussion}
In this section, we explore the broader implications of our work, potential extensions, and future directions.
We discuss compensation mechanisms, paths toward full decentralization, system limitations, alternative deployment models, and possible generalizations.

\subsection{Deployment Variants}
Community proving can be deployed in various ways, depending on how the rollup operator handles batch proof computation. Here, we discuss two primary deployment models: \emph{In-house proving} and a \emph{Prover marketplaces}.

\point{In-House Proving}
In the first model, the batch proof is computed directly by the rollup operator, as is currently the case with ZKsync. The operator may choose to outsource specific proving tasks to the community while continuing to handle others in-house. In ZKsync's current setup, proving machines (virtual machines on Google Cloud) access a database of available proving jobs, complete the jobs, and then submit the results back to the database.

In this model, the Job Distributor~(JD) can be implemented as a special prover machine that interacts with the system in the same way as existing cloud-based provers. However, instead of performing the job locally, the JD outsources the computation to a community prover. This design is straightforward to implement as it requires minimal changes to the core Boojum infrastructure. The interface exposed by the JD is identical to that of a regular prover, ensuring seamless integration with the existing system.

This approach also naturally mitigates any liveness concerns related to community proving. By default, all jobs are handled by the rollup operator, with community provers taking on as many jobs as they are able to. In ZKsync’s Boojum deployment, the cloud-based commissioning of prover virtual machines is already dynamic, scaling according to the real-time demand. Therefore, the addition of community provers automatically reduces the need for cloud provers, offering significant cost savings while maintaining system stability.



\point{Prover Markets}
In a second model, the rollup operator selects proof providers from a decentralized marketplace. Here, community proving can be implemented as a ``prover pool'' that functions as a single entity in the marketplace. The pool operator deploys an instance of Boojum, which then outsources the individual proving jobs to community provers. The rewards earned from the marketplace are shared among participating community provers, incentivizing their involvement.

This model also supports partial outsourcing, similar to the in-house approach, but with one key distinction: instead of the rollup operator managing the process, the pool operator takes on this role. Community provers are thus pooled together under a single entity that interacts with the marketplace, allowing for a more decentralized and distributed proof generation process.


\subsection{Smart Job Assignment}
One promising direction for future improvement is enhancing the Job Distributor~(JD) by incorporating smarter job assignment algorithms that account for the reliability of provers. Currently, jobs are assigned to available provers, but the system could be improved by evaluating each prover's performance history, availability, and hardware capabilities. For example, provers with a track record of high reliability~----~such as consistently completing jobs within the expected time frame and producing valid proofs~---~could be prioritized for more computationally demanding tasks. Additionally, assigning jobs based on hardware type, such as favoring GPU-based provers for tasks that benefit from parallelization, would optimize resource utilization and further reduce proving time.

Another area of improvement is job reassignment. When jobs need to be reassigned due to a failed or delayed response from a prover, there is a risk that the same prover may be reassigned the same job multiple times. To mitigate this, the system could implement a strategy that prioritizes provers differently based on their historical performance and current availability. This approach would ensure a fair distribution of workload among provers, while avoiding overburdening unreliable or overloaded provers. By intelligently selecting a diverse set of provers for reassignment, the system can better balance the load and reduce the risk of bottlenecks caused by repeated job failures.

\subsection{Prover Compensation}
While the exact mechanisms for compensating community provers are beyond the scope of this paper, several models can be considered, each offering different trade-offs in terms of cost, performance, and decentralization.

\point{Fixed-Rate Compensation}
\label{fixed_rate_compensation}
A straightforward approach is a fixed-rate compensation model, where each prover is paid a predetermined amount in cryptocurrency (or fiat currency) for every successful proof. This model is well-suited for rollups controlled by a central operator, as is still the case for many of today's popular ZK rollups. Under this scheme, the provers submit their wallet address along with the proof, and upon validation, the job distributor transfers the reward to the attached address.

Although this model offers predictable and stable costs, it may not appropriately incentivize high-performance provers. More powerful machines equipped with GPUs would receive the same compensation as less capable hardware, potentially leading to sub-optimal resource utilization. Despite its simplicity, this model might fail to encourage provers to invest in faster hardware, limiting the system's overall efficiency.

\point{Time-Based Compensation}
Alternatively, a time-based compensation model better aligns incentives with the performance of the prover. In this model, faster provers, such as those using GPUs, are rewarded more for their quicker job completion times. This results in reduced proving times but can also lead to higher costs. However, in latency-sensitive scenarios, where the cost of delay outweighs the cost of higher compensation, this approach could be particularly beneficial. By rewarding efficiency, this model encourages provers to invest in better hardware, ultimately improving system performance while balancing the cost trade-off.

\point{Adaptive Pricing Models}
To create a more dynamic and responsive system, adaptive pricing models could be introduced, allowing compensation to fluctuate based on demand and prover availability. For example, during periods of high demand, the reward for successfully completing a proof would increase, incentivizing more provers to participate. In contrast, during periods of low demand, the compensation could decrease to conserve resources. This model would create a flexible ecosystem where provers are incentivized to contribute when their services are most needed, leading to improved efficiency and optimized cost management.

\point{Volunteer Labor}
In a more decentralized rollup, particularly one serving a community-driven network, it is possible that some users may voluntarily execute proving jobs without any compensation. These users could contribute idle computational resources that would otherwise go unused. While this approach is more speculative, it offers a vision of a community-powered rollup where decentralization is achieved not just through technical means, but also through voluntary participation and shared responsibility.

\point{Marketplace-Based Compensation}
A fine-grained marketplace model could be implemented, similar to existing marketplaces for full batch proofs, but focused on individual jobs or smaller batches. In this setup, the provers could bid for jobs in an auction mechanism; either a sealed-bid auction, where provers bid for specific tasks, or a Dutch reverse auction, where the jobs are awarded based on who can complete them the fastest. This marketplace model is more complex than fixed or time-based compensation, but offers greater flexibility. Since individual jobs are smaller and faster to compute, the system could handle failures more easily, making this a viable approach to scaling the proving process while incentivizing provers through competitive bidding.

\subsection{Economic Incentives and Security}
To further enhance the security and integrity of the community proving system, one potential improvement is the introduction of cryptographic bonds for provers. In this model, the provers would be required to lock up a certain amount of collateral (a bond) before participating in the proving process. This bond acts as a financial stake, ensuring that the provers have an incentive to behave honestly.

If a prover is found to submit incorrect or malicious proofs, they would forfeit part or all of their bond as a penalty. This economic disincentive serves as a strong deterrent against dishonest behavior, as provers are unlikely to risk losing their collateral. By implementing such a mechanism, the system ensures that only provers with a genuine interest in contributing honestly and efficiently will participate, ultimately improving the reliability of the network.

\subsection{Complete Decentralization}
As discussed in Section~\ref{sec_intro}, our current approach to community proving in \tool remains technically centralized, as the rollup operator retains full control over the proof construction process. The coordination of community provers lacks transparency, and the orchestration infrastructure still presents a single point of failure.

To move towards fuller decentralization, the coordination mechanism itself could be replicated and modeled as a state machine, executed by a decentralized system. It is not necessary to develop a dedicated decentralized platform specifically for this purpose, although that is an option. Instead, existing blockchain systems with sufficiently expressive smart contract platforms can be used to manage the coordination, with some platforms being more suitable than others depending on their throughput and latency. Ideally, the underlying system should provide high throughput and low latency, as provers will need to interact with it frequently.

An interesting possibility is using the rollup system itself to coordinate the community proving process. In this scenario, the state of the proving orchestration logic observed by the provers would not be finalized, as it would be part of the very process of generating the proofs. While this might seem counterintuitive, it simply means that provers would be working with the unfinalized state of the orchestrator. The main downside of this approach is that, in the event of a rollback (or ``reorg'') of unproven rollup states, some proving work could be lost. However, given that such reorgs are very rare in practice, the impact of this risk is minimal.
\section{Related Work}
\label{sec:related}
The growing demand for scalability in zero-knowledge~(ZK) proof systems has led to significant advancements in parallelization, distributed proving, and community-based computational models. In this section, we explore some key projects that align with our approach to community proving, focusing on efforts to enhance ZK proving efficiency, decentralize the proving process, and leverage volunteer-based computing networks for decentralized infrastructure.

\subsection{Parallelizing ZK Proving}
Prior work has made significant strides in an effort to parallelize existing proof systems and run them on clusters of trusted and reliable machines as opposed to single computers, primarily to shorten proof generation time. However, this does not advance the goals of decentralization. 
In contrast, \tool focuses on taking tasks that can be obtained from pre-existing distribution strategies from existing rollups and their default provers, including ZKSync or Polygon's zkEVM, or research systems such as DIZK~\cite{wuDIZKDistributedZero} and Pianist~\cite{liuPianistScalableZkRollups}. As such, \tool takes tasks produced by these systems as black boxes and provides an orchestration layer that runs on untrusted hardware with permissionless participation.

\point{Distributing ZK proving}
DIZK is a pioneering system that decentralizes ZK proof generation using multiple machines in a compute cluster. By employing novel methods for distributing proof tasks across machines, DIZK scales to handle computations involving billions of logical gates~---~100 times larger than previous efforts~---~at a processing cost of approximately~10 microseconds per gate, representing a significant performance improvement over some earlier systems. This capability opens the door for ZK proofs to be applied to more complex security applications, demonstrating the potential of parallelism in distributed ZK proof generation~\cite{wuDIZKDistributedZero}.

Further advancing the scalability of ZK rollups, Liu et al. propose Pianist, a fully distributed system for generating ZK proofs across multiple machines. Based on the Plonk proof system, Pianist enables provers to split the proving task among many machines with minimal inter-machine communication, significantly improving the scalability of Plonk. The authors demonstrate that Pianist can generate proofs for~8,192 transactions in just~313 seconds using~64 machines, achieving a~64-fold improvement in scalability. Note that the machines used in these experiments are quite powerful, using~256~GB of RAM, largely precluding broader participation. This system offers a scalable approach to ZK rollups, where distributed proof generation can be performed efficiently without compromising performance~\cite{liuPianistScalableZkRollups}.

Building on Pianist, Li et al. introduce HyperPianist, a distributed proof system characterized by linear-time prover complexity and logarithmic communication overhead. The system leverages a distributed SumCheck protocol, achieving logarithmic communication costs among distributed machines. This model offers moderate improvements in proving efficiency, ranging from $1.41\times$ to $1.89\times$ on small inputs, showing promise for further scaling ZK proofs in decentralized environments~\cite{liHyperPianistPianistLinearTime}.

\point{New Frameworks and Languages}
To further streamline the ZK proof generation process, Nguyen et al. introduce Mangrove, a framework for generating folding-based SNARKs. Mangrove's ``uniformizing'' compiler transforms any NP statement into a series of identical steps, allowing the folding-based incremental verification~(IVC) scheme to be more efficient. The framework improves upon traditional recursive proof systems by reducing overhead and optimizing the memory usage of provers. These innovations result in a prover system that requires only two passes over the data, offers highly parallelizable computation, and is both memory-efficient and concretely performant, rivaling some top monolithic SNARKs~\cite{nguyenMangroveScalableFramework2024}.

In the realm of programming languages for ZK proofs, Sang et al. introduce Ou, a novel language designed to automate the parallelization of ZK proofs. Coupled with the Lian compiler, which automates the analysis and partitioning of program statements across a computing cluster, Ou simplifies the process of writing efficient ZK proofs. The system uses formal methods and combinatorial optimization to split a ZK program into optimized segments, facilitating parallel execution and verification~\cite{sangOuAutomatingParallelization2023}. This approach offers a promising step forward 
for ZK-proof developers, particularly for scaling the proving process across distributed machines.

Hu et al.~\cite{huParallelZeroknowledgeVirtual2024} present a zkVM design that uses data-parallel circuits to parallelize operations at both the opcode and basic block levels. This design reduces the overhead traditionally associated with zero-knowledge circuits by dynamically adjusting the circuit size based on actual opcode usage. While promising in its ability to optimize performance for zkVMs, this work lacks empirical comparisons to other approaches, making it challenging to evaluate its real-world impact.

\subsection{Volunteer-Based Compute}

The concept of volunteer-based computing, where participants contribute idle computational resources to perform large-scale distributed tasks, has been successfully deployed in various contexts, including scientific projects such as SETI@Home and Folding@Home.
In the blockchain space, similar efforts have been made to leverage community resources for decentralized computing tasks.
Projects such as Golem\footnote{\url{www.golem.network}} and Filecoin\footnote{\url{filecoin.io}} incentivize participants to contribute storage or computational power to the network in exchange for rewards, enabling decentralized storage and compute networks to flourish.

More recently, community-based proving networks have emerged, connecting idle computational resources with projects requiring capacity for ZK proof generation. Companies like Gevulot\footnote{\url{www.gevulot.com}} have created platforms where community members can contribute their computational resources to participate in the proving process. These volunteer-driven networks hold promise for further decentralizing the proving infrastructure, reducing costs for operators while engaging a wider community of participants in the ZK rollup ecosystem.
\section{Conclusion}
\label{sec_conclusion}
In this paper, we introduce \tool, a novel approach to decentralizing the computationally intensive task of generating validity proofs for ZK rollups, leading to what we call \emph{community proving}. By offloading selected proving tasks to community participants, our approach achieves a balance between decentralization and performance, reducing infrastructure costs for rollup operators, while offering financial incentives for public prover participants to contribute their computational resources.

We designed and implemented \tool for the Boojum prover used in ZKsync, demonstrating through empirical evaluation that Community Proving can deliver competitive performance compared to centralized cloud-based proving setups. Our results show that even with consumer-grade hardware, community provers can match, and in some cases surpass, the efficiency of centralized deployments. 
For instance, our experiments demonstrate that XXX...
This opens the door to broader participation in the ZK rollup ecosystems, fostering greater community involvement and reducing the reliance on expensive cloud infrastructure, creating a natural path of decentralization.

Although our approach brings clear benefits in terms of cost savings and community engagement, it is important to recognize that full decentralization has not yet been achieved. At present, the orchestration of the proving process in \tool remains centralized, representing a key area for future research and development. Future work could explore decentralized coordination mechanisms, cryptographic bonds for provers to ensure reliability, and adaptive compensation models.

Overall, our work lays the foundation for a more inclusive and scalable proving architecture in ZK rollups, moving toward a future where proof generation is distributed across a diverse network of decentralized contributors, thereby enhancing both the security and efficiency of blockchain systems.

\bibliographystyle{ACM-Reference-Format}
\bibliography{sample-base}

\end{document}